\documentclass[11pt]{article}
\usepackage{graphicx,caption2,subfigure}
\makeatletter
\newcommand{\singlespacing}{\let\CS=\@currsize\renewcommand{\baselinestretch}{1.0}\tiny\CS}
\newcommand{\doublespacing}{\let\CS=\@currsize\renewcommand{\baselinestretch}{1.5}\tiny\CS}

\begin{document}
\title {On the Special Significance of the Latest PAMELA Results in Astroparticle Physics}
\author { Goutam
Sau$^1$\thanks{e-mail:sau$\_$goutam@yahoo.com}$\&$ S.
Bhattacharyya$^2$\thanks{e-mail: bsubrata@www.isical.ac.in
(Communicating Author).}\\
{\small $^1$ Beramara RamChandrapur High School,}\\
 {\small South 24-Pgs,743609(WB),India.}\\
 {\small $^2$ Physics
and Applied Mathematics Unit(PAMU),}\\
 {\small Indian Statistical Institute, Kolkata - 700108, India.}}
\date{}
\maketitle
\bigskip
\begin{abstract}In continuation of their earlier measurements, the
PAMELA group reported data on antiproton flux and $\overline{P}/P$
ratios in 2010 at much higher energies. In past we had dealt with
these specific aspects of PAMELA data in great detail and each time
we captured the contemporary data-trends quite successfully with the
help of a multiple production model of secondary antiprotons with
some non-standard ilk and with some other absolutely standard
assumptions and approximations. In this work we aim at presenting a
comprehensive and valid description of all the available data on
antiproton flux and the nature of $\overline{P}/P$ ratios at the
highest energies reported so far by the PAMELA experiment in 2010.
The main physical implication of all this would, in the end, be
highlighted.
 Cosmic rays (including sources, origin, acceleration, and interactions).
 Dark Matter (stellar, interstellar, galactic, and cosmological). 
\end{abstract}
\newpage
In the cosmic ray physics and astroparticle domain the most
familiar and important antiparticles are the positrons and
antiprotons. The energy dependent behaviours of $e^+/e^-$ and
$\overline{P}/P$ are experimentally found to be sharply in
contrast. While for the former (positron-to-electron ratio) the
antiparticle-to-particle ratio experimentally shows a slowly
rising trend with increasing energy, the behaviour for the latter
is seem to be quite different. And the controversy persists over
the question whether $\overline{P}/P$ ratios remain nearly
constant with increasing antiproton energy or there is a gradual
fall-off of the ratio with a not-very-sharp slope. Besides, it is
well known that cosmic ray antiproton energy spectrum provides
valuable information on the origin and propagation of cosmic rays.
It also throws light on the feasibility of the `exotic' sources of
primary antiprotons, such as annihilations of dark matter
particles and the evaporation of the primordial black holes.
\par
Our work here would be grounded on the following physical
considerations which are being listed at the outset. (i) We would
counter the hypothesis of the exotic sources of antiprotons, nor we
would assume any contribution from the supersymmetric
particles-decays. (ii) We accept the simple leaky box model for
propagation of cosmic rays. (iii) We will assume here the primary
proton spectra represented by Bhadwar et al\cite{Bhadwar1} to remain
valid. (iv) We concentrate only on the secondary antiprotons which
arise of the multiparticle production phenomena at high/superhigh
energies. (v) We induct a multiple production model which is
certainly not of the typically `standard' variety. In fact, herein
an element of exoticity would come into the whole of the physics
scenario. (vi) As the present study pertains to much higher
energies, the probability of occurrence of contributions from
annihilation channels would be considered to be totally switched off
in our mathematical calculations, for which the damping
term\cite{Sau1} that was introduced in one of our previous works on
the same topic would be eliminated here altogether.
\par
According to the model the low-$p_T$ (soft) baryon-antibaryons are
produced through the decays of (virtual) secondary pions in a
sequential chain of which proton-antiproton pairs comprise nearly
one third of the total. Bhattacharyya\cite{Bhattacharyya3} had
worked out the details of the necessary field-theoretic calculations
based on Feynman diagrams and obtained the following formulae for
inclusive cross-sections at low-$p_T$ valid for moderately high to
very high energies and the expression for average antiproton
multiplicity
\begin{equation}
E\frac{d^3\sigma}{dp^3}|_{pp\longrightarrow\overline{p}X}\simeq1.87 \times exp[-7.38\frac{p{_T}^{2}+m^{2}_{\overline{p}}}{1-x}]exp[-5.08x]
\end{equation}
and
\begin{equation}
<n_{\overline{p}}>\simeq1.08 \times
10^{-2}s^{2/5}~~~~~~~~~~~~~~~~~~~~~~~~~~~~~~~ for
\sqrt{s}\leq100GeV
\end{equation}
\begin{equation}
<n_{\overline{p}}>\simeq2 \times
10^{-2}s^{1/4}~~~~~~~~~~~~~~~~~~~~~~~~~~~~~~~ for \sqrt{s}>100GeV
\end{equation}
where $m_{\overline{p}}$ is the mass of the antiproton and
$n_{\overline{p}}$ is the measured antiproton multiplicity. With
(2) we get at $\sqrt{s}$=53GeV, $<n_{\overline{p}}>\simeq$0.2 for
both the formulae.
\par
In actual evaluation of the cosmic antiproton production
observables the model dependence comes into picture for getting
values of d${\overline{\sigma}}$/dE which is related with the
inclusive cross-section in the following way\cite{Tan1} :
\begin{equation}
\frac{d\overline{\sigma}}{dE}=\frac{\pi}{p_L}\int(E\frac{d^3\sigma}{dp^3})_{pp\rightarrow\overline{p}X} ~ dp_T^2
\end{equation}
Here we take p$_L\simeq$E as the transverse momenta of the produced
secondary antiprotons is assumed to be small.
\par Inserting expressions (1), (2) and (3), our model-derived
formula for inclusive cross-section valid at moderately high
energies in eqn.(4) and integrating over $p_T$ with normal
approximations we get
\begin{equation}
\frac{d\overline{\sigma}}{dE}|_{p\rightarrow\overline{p}}\simeq0.496~exp[-5.08x]
\end{equation}
where we have used the low-transverse-momentum upper limit up to
$p_T$= 1 GeV/c. It must also be recalled that $p_L\simeq$E.
Bhattacharyya and Pal\cite{Bhattacharyya1,Bhattacharyya2} have
worked out that the antiproton-to-proton ratio is to be given
finally by
\begin{equation}
f_{\overline{p}}(E)=\frac{J_{\overline{p}}(E)}{J_p(E)}=\frac{2K\lambda_e(E)}{m_p}\int^{X_s}_0E\frac{d\sigma_p}{dE}X^{\gamma-1}dx
\end{equation}
where $J_p$ and $J_{\overline{p}}$ are the differential fluxes of
the primary protons and the secondary antiprotons ($m^{-2} sr^{-2}
s^{-1} GeV^{-1}$) respectively. K is the correction coefficient
taking into consideration the composition of the primary cosmic rays
and the interstellar gas, $\lambda_e(E)$ is the average path length
of antiprotons against escape ($g cm^{-2}$ as the unit), $m_p$ is
the mass of the proton (g as the unit), $E_p$ is the total energy of
the primary proton, $E_s$ is the integral lower limit relevant to
the production threshold of antiprotons, $\gamma$ is the integral
energy spectrum exponential of the primary protons and is the sole
quantity taken from cosmic-ray information. For actual calculations
we use here : $\gamma$=1.75, X = E/$E_p$ and $X_s$ = $E_s/E_p$ (we
took $X_s \simeq$ -($m_pc^2$/E)+[$(m_pc^2/E)^2$+1]$^{1/2}$).
\par Usually, $J_p$ is expressed as
\begin{equation}
J_p(E_p) = J_0 E_p^{-(\gamma+1)}
\end{equation}
Now using eqn.(5), eqn.(7) in eqn.(6) we get
\begin{equation}
\frac{f_{\overline{p}}(E)}{K\lambda_{e}(E)}=\frac{2}{m_p}\int_0^{X_s}0.496 ~ exp[-5.08x]x^{\gamma-1}dx
\end{equation}
Here, we have always used K = 1.26, $\lambda_e$ = 5 $g cm^{-2}$ and
$m_p \simeq$ 1GeV.
\par
\par In figure 1, the median energy of primary protons is shown
as a function of the secondary antiproton energy. It was found
earlier that for antiproton energies of 3-9 GeV, which were relevant
to the experimental work performed uptil then, the median energies
of primary protons were about 25-80 GeV. The present energy-region
is much higher. But we take the cue from Tan and Ng\cite{Tan1} and
proceed in a similar manner to draw the figure shown in figure 1 as
described in the figure-caption in some detail. Very carefully, we
have chosen a modestly accurate primary proton spectrum. Modifying
Bhadwar et al\cite{Bhadwar1} , we use here
\begin{equation}
J_P(E_P) = 2 \times 10^5 E_P^{-2.75}
\end{equation}
where $J_P(E_P)$ is in protons $m^{-2} sr^{-2} s^{-1} GeV^{-1}$.
\par
The final results have here been actually worked out on the basis of
the following two deduced expressions :
\begin{equation}
f'_{\overline{P}}(E_{\overline{P}}) = f_{\overline{p}}(E) J_P(E_P)
\end{equation}
and
\begin{equation}
R_{\overline{p}}(E_{\overline{P}})=\frac{J_{\overline{p}}(E)}{J_p(E)} =
 \frac{2K\lambda_e(E)}{m_p}\int^{X_s}_0E\frac{d\sigma_p}{dE}X^{\gamma-1}dx
\end{equation}
The graphical plots presented in figure 2 and figure 3 are done with
the help of eqn.(8). The expression (10) describes the nature of
data measured by PAMELA group and others on antiproton production
flux. And the plot of model-based $\overline{P}/P$ ratio-values
based on expression (11) are displayed in figure 4 and figure 5
against the data-background. The used values of the parameters are
shown in the adjoining table (Table 1).
\par
Very recently, there has been a new twist in the situation vis-a-vis
the $\overline{P}/P$ ratio behaviour with the availability of the
results on the same observable ($\overline{P}/P$ ratios) measured by
ARGO-YBJ collaboration\cite{Sciascio1}. The measured values depict
the $\overline{P}/P$ values at some higher ranges of values than
what could be expected of just an extrapolation of the
PAMELA-data-trends.
\par
Some comments are in order for reported data by ARGO-YBJ (AY)
Collaboration, especially because of the fact that their
measurements put the $\overline{P}/P$ ratios apparently to much
higher values. The word `apparent' used here is firstly to draw
attention to the large error bars indicated by this AY Collaboration
through the downward-oriented arrow marks. The energy values are
much higher for AY Collaboration - so are the ranges of
uncertainties in measurement. But in the literature related to the
studies at TeV energy region the terms, like, primary and secondary,
seem to have been messed up. The definition of `primary' used by
Preghenella\cite{Preghenella1} is at variance with what the
connotation of the word is in Cosmic Ray Physics. Still, there is a
striking commonness between the tendencies of the measured data to
measure and report the $\overline{P}/P$ ratios quite at high values.
This is somewhat possible in purely nuclear collisions and also in
nucleus-dominated cosmic interactions\cite{Adriani2} at very high
energies. Still, such high values of $\overline{P}/P$ ratios seem
unlikely, for which further scrutiny by the experiments is surely
warranted.
\par
And this newest piece of observation or evidence pushes the
controversy on $\overline{P}/P$ nature to a different pitch. The
predicted rising nature of $\overline{P}/P$ ratio with
$\overline{P}$ energy is a new addition. Thus the possibilities that
arise are threefold : (i) nearly constant (or steady) nature of
$\overline{P}/P$ ratio (ii) decreasing value of the ratio and (iii)
the rising ratio of $\overline{P}/P$.
\par
There is another not-very-clear and questionable aspect in ARGO-YBJ
experiment. The measurement or the Monte Carlo Simulations were
apparently made for matter + antiproton production. Firstly, no
specific comment on what `matter' represents here is given by the
Collaboration. Secondly, the protonic content was taken to be nearly
72\% on the average on an absolutely arbitrary basis. The particular
model which obtains the $\overline{P}/P$ values at such high orders
is also based on the assumption of emission of primary antiproton
production by antistars. This is simply too speculative; the other
models which depict similar high $\overline{P}/P$ values referred to
by ARGO-YBJ Collaborations are also based on some wild assumptions;
and so they are under the spell of strong doubts. Thus, on an
overall and general basis, we are not in favour of attaching too
much importance to ARGO-YBJ results at this point of time.
\par
It is to be noted that PAMELA results, especially those reported in
2010, cover substantially quite a high energy band. We are of the
opinion that it would probably be more justified to use the nature
of the primary proton spectrum suggested by JACEE
Collaboration\cite{Asakimori1,Burnett1}. This would surely be given
a fair trial in one of our future works.
\par
The implication of this work will now be spelt out. Contributions
arising out of the exotic sources, like, (i) dark matter particles,
dark matter annihilations etc., (ii) supernova remnants and (iii)
emission from the pulsars are being ruled out in our approach to
this work. Besides, matter-antimatter components expected to be
emanated from any extragalactic sources are also entirely excluded
in our calculations. We also assume that the solar modulation
phenomenon affects both proton and antiproton in an identical
manner. So, it does not have any effect, at least on
$\overline{P}/P$ ratios. Finally, the emphasis rests mainly and only
on the choice of a heretic multiple production model. So, the focus
of the problem is being transferred here from the domain of
Astroparticle Physics to the sphere of Particle Physics. In essence,
this represents virtually a case of `Paradigm Shift' in the area of
contemporary Astroparticle Physics.

\singlespacing

\newpage
\begin{table}
\begin{center}
\begin{small}
\caption{Chosen values of the parameters}
\begin{tabular}{|c|c|c|c|}\hline
$\lambda_e(g cm^{-2})$ & $m_p(GeV)$ & $\gamma$ & $K$ \\
\hline
$1.26$&$\simeq1.00$& $1.75$ & $5.00$\\
\hline
\end{tabular}
\end{small}
\end{center}
\end{table}

\begin{figure}
\centering
\includegraphics[width=2.5in]{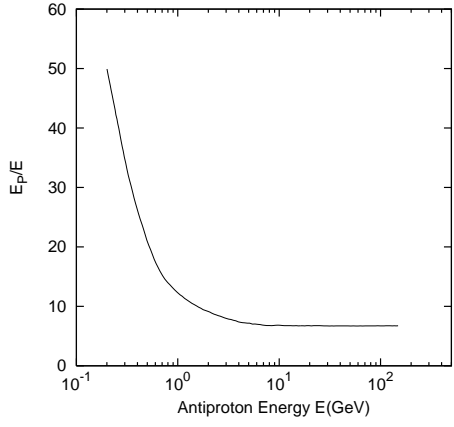}
\caption{Relation between primary proton energy ($E_P$) and the total antiproton energies (E);
 the plot has been done with $E_P/E$ as Y-axis and E-valus as X-axis.}
\end{figure}

\begin{figure}
\centering
\includegraphics[width=2.5in]{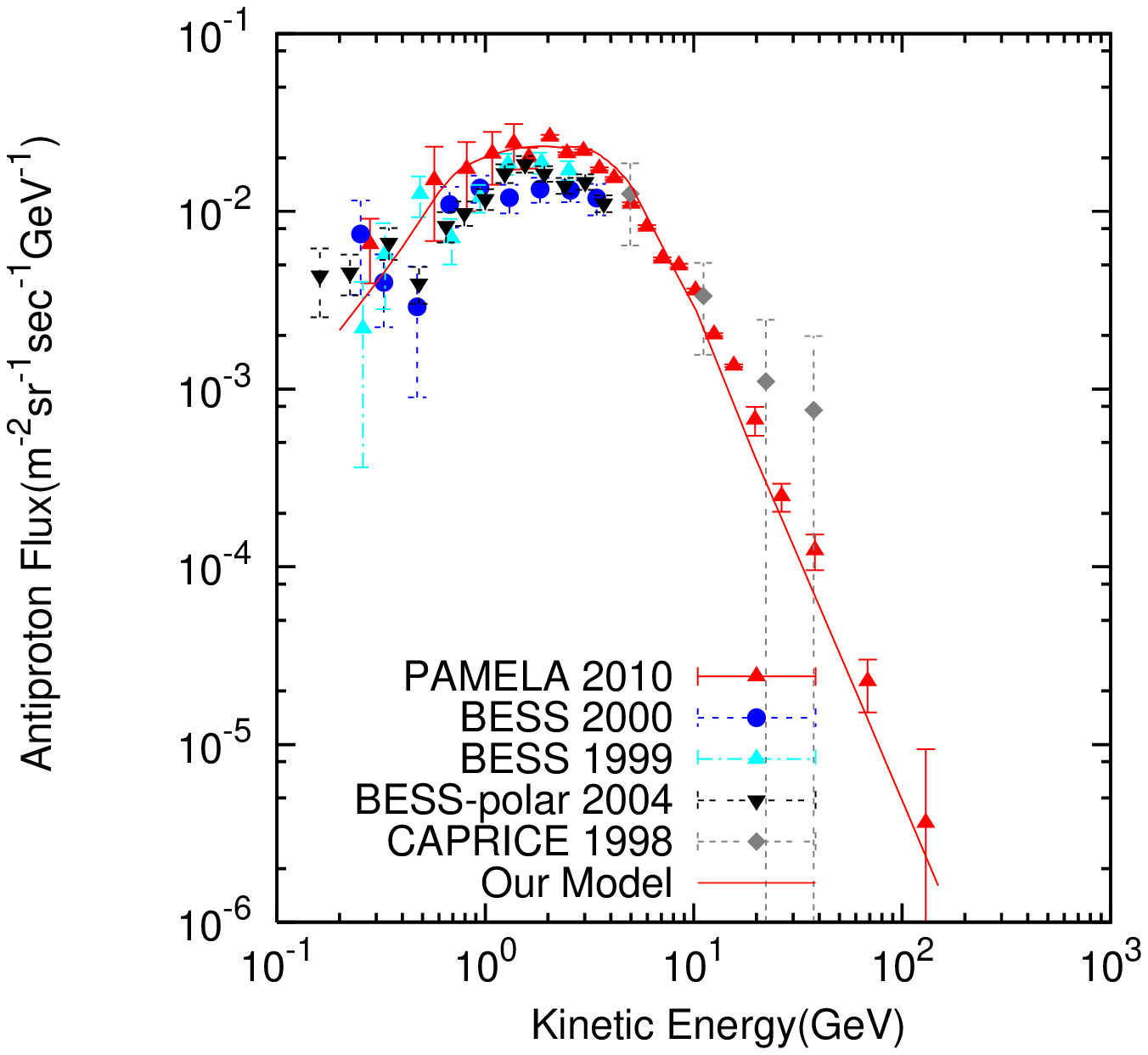}
\caption{The antiproton energy spectrum at the top of the payload as measured by the PAMELA group and others.
The solid curves shows the calculations of our theoretical model (10).
The experimental data are collected from Ref.\cite{Adriani1}-\cite{Abe1} }
\end{figure}

\begin{figure}
\centering
\includegraphics[width=2.5in]{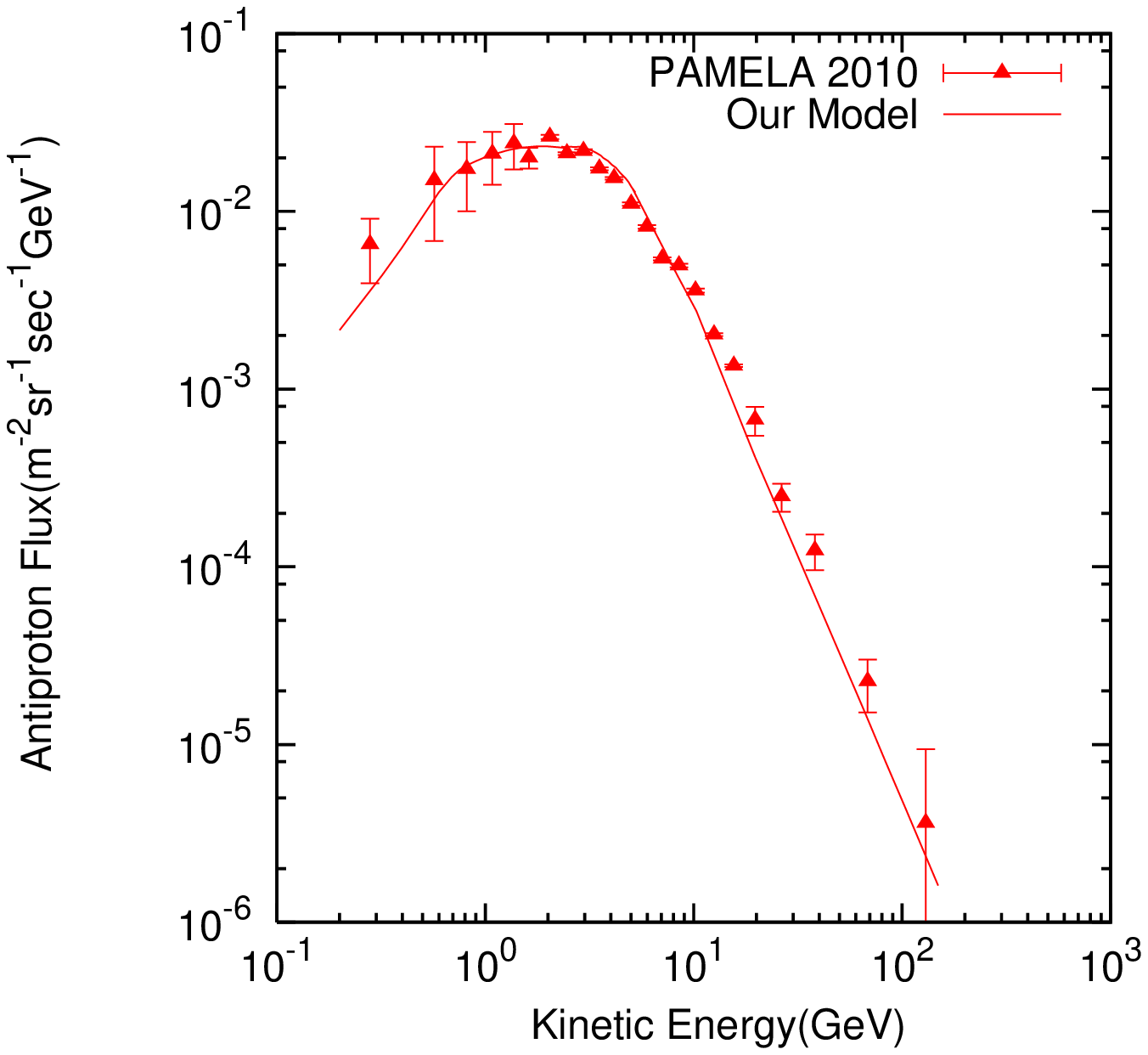}
\caption{The antiproton energy spectrum at the top of the payload as measured by the PAMELA group.
The solid curves shows the calculations
of our theoretical model (10). The experimental data are collected from Ref.\cite{Adriani1} }
\end{figure}

\begin{figure}
\subfigure[]{
\begin{minipage}{.5\textwidth}
\centering
\includegraphics[width=3.0in]{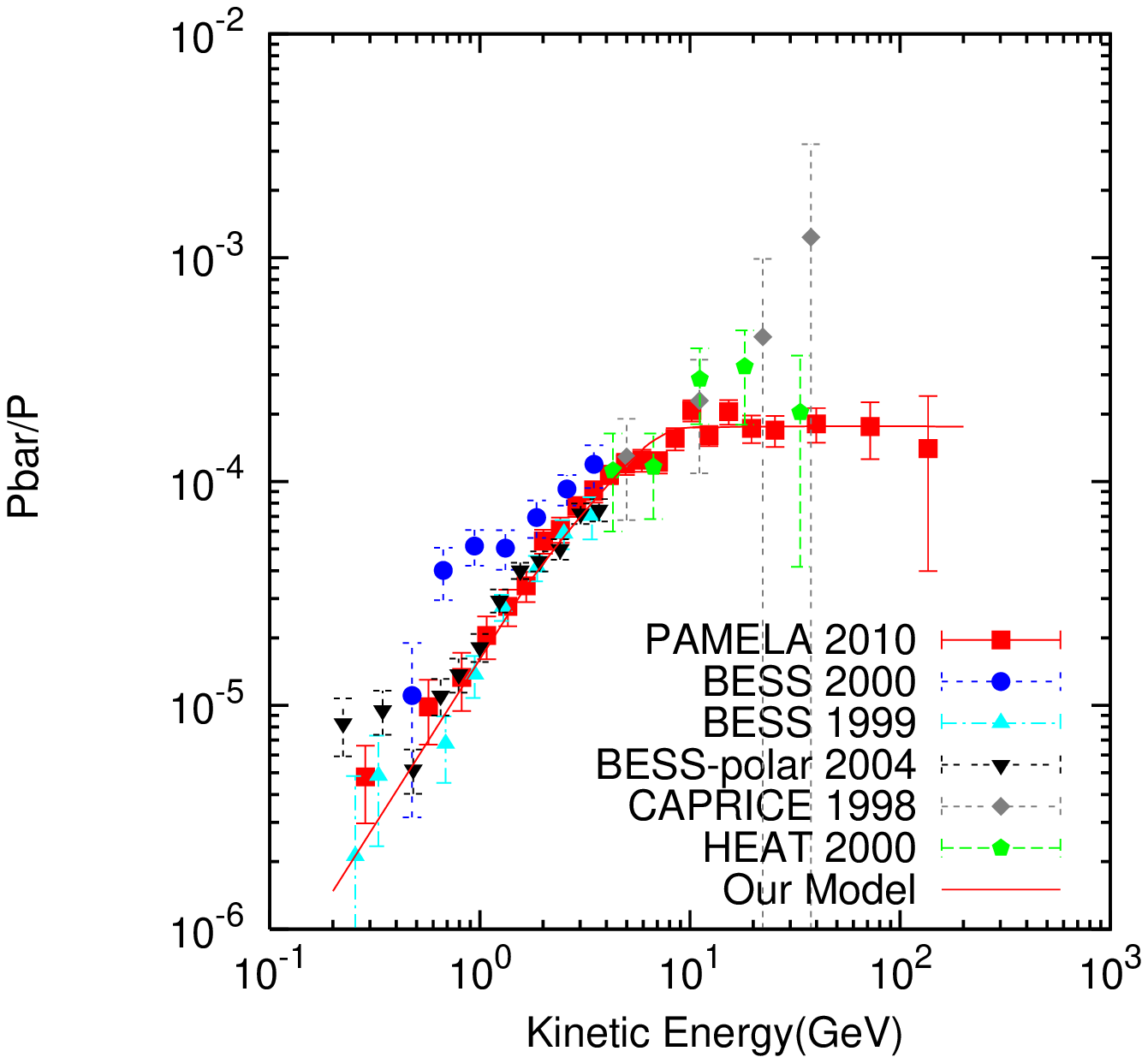}
\end{minipage}}%
\subfigure[]{
\begin{minipage}{.5\textwidth}
\centering
 \includegraphics[width=3.0in]{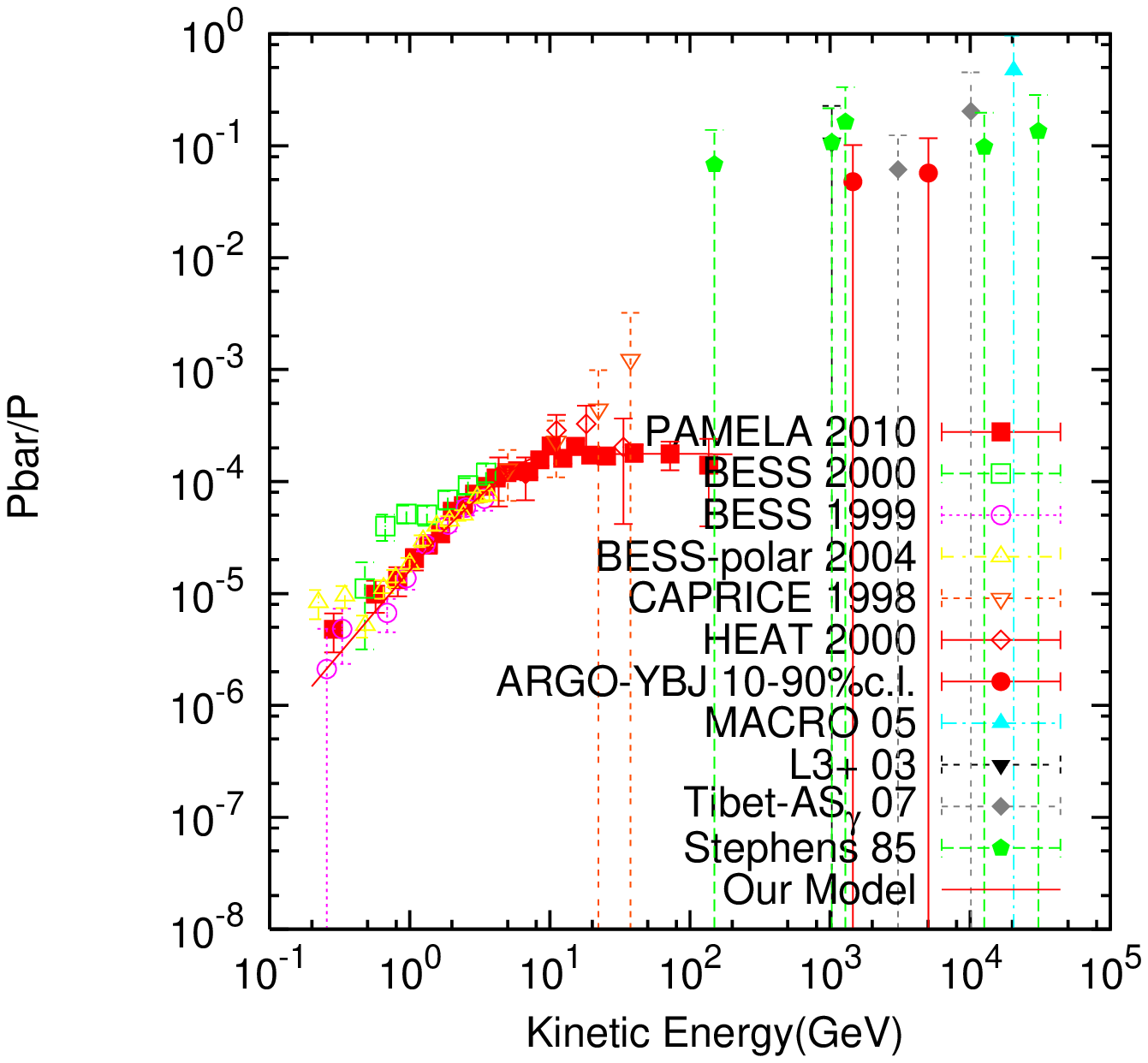}
 \end{minipage}}%
\caption{The antiproton-to-proton flux ratio at the top of the payload as measured by PAMELA group and others.
The solid curves represent our calculations based on our model based approach (11).
 The experimental data are collected from Ref.\cite{Sciascio1}, \cite{Adriani1}-\cite{Beach1} }
\end{figure}

\begin{figure}
\centering
\includegraphics[width=2.5in]{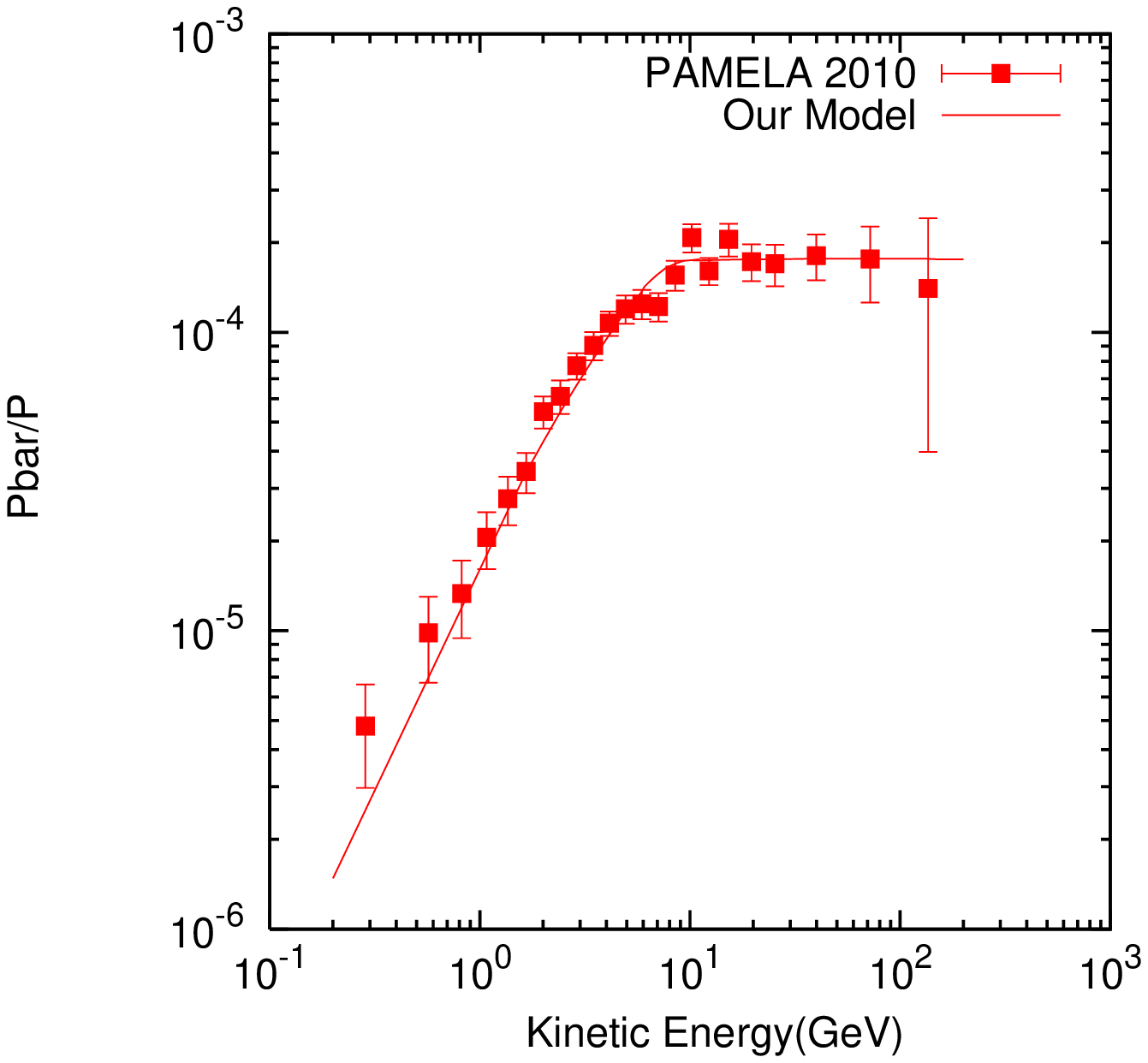}
\caption{The antiproton-to-proton flux ratio at the top of the payload as measured by PAMELA group
and the solid curves represent our calculations based
on our model based approach (11). The experimental data are collected from Ref.\cite{Adriani1}}
\end{figure}

\end{document}